\documentclass[
plb,
superscriptaddress,
tightenlines,
showpacs,
showkeys,
nofootinbib,
aps,
amsfonts,
amssymb,
twocolumn
]{revtex4}

\usepackage[usenames,dvipsnames]{xcolor}
\usepackage{amsmath}
\usepackage{amsfonts}
\usepackage{amssymb} 
\usepackage{graphicx}
\usepackage{epic}
\usepackage{eepic}
\usepackage{epsfig}
\usepackage{latexsym}
\usepackage{float}
\usepackage{shorthand}
\usepackage{multirow,array}
\usepackage[caption=false]{subfig}
\PassOptionsToPackage{hyphens}{url}

\usepackage[paperwidth=210mm,paperheight=297mm,centering,hmargin=1.7cm,vmargin=2.1cm,bindingoffset=0.2cm]{geometry}

\usepackage[
    colorlinks,
    linkcolor=green!40!blue,           
    citecolor=red,            
    filecolor=green!40!blue,        
    urlcolor=green!40!blue,         
    hyperfootnotes]{hyperref}

\renewcommand{\lq}{\ell_q}
\renewcommand{\lg}{\ell_8}
\newcommand{\ubr}[1]{\raisebox{1.5ex}{\hspace{#1ex}$\frown$\relax}}
\newcommand{\lbr}[1]{\raisebox{-1.5ex}{\hspace{#1ex}$\smile$\relax}}

\begin{document}

\title{{\hfill{\rm\small MITP/16-012}\\}Probing Compositeness with the CMS $eejj$ \& $eej$ Data}

\author{Tanumoy Mandal}
\email{tanumoy.mandal@physics.uu.se}
\affiliation{Department of Physics and Astronomy, Uppsala University,
Box 516, SE–751 20 Uppsala, Sweden}

\author{Subhadip Mitra}
\email{subhadip.mitra@iiit.ac.in}
\affiliation{Center for Computational Natural Sciences and Bioinformatics,
International Institute of Information Technology, Hyderabad 500 032, India}

\author{Satyajit Seth}
\email{sseth@uni-mainz.de}
\affiliation{PRISMA Cluster of Excellence, Institut f\"{u}r Physik,
Johannes Gutenberg-Universit\"{a}t Mainz, 
D\,-\,55099 Mainz, Germany}


\begin{abstract}
Quark-lepton compositeness is a well-known beyond the Standard Model (SM) scenario with heavy exotic particles like leptoquarks (LQs) and leptogluons (LGs) etc. These particles can couple to leptons and jets simultaneously. In this letter, we use the recent CMS scalar LQ search data in the $eejj$ and $eej$ channels to probe this scenario. We recast the data in terms of a color octet partner of the SM electron (or a first generation spin-1/2 LG) that couples to an electron and a gluon via a dimension five operator suppressed by the quark-lepton compositeness scale ($\Lm$). By combining different production processes of the color octet electron ($e_8$) at the LHC, we use the CMS 8TeV data to obtain a simultaneous bound on $\Lm$ and the mass of the $e_8$ ($M_{e_8}$). We also study the reach of the 13 TeV LHC to discover the $e_8$ and interpret the required luminosity in terms of $M_{e_8}$ and $\Lm$.
\end{abstract}



\maketitle

\section{Introduction}\label{sec:intro}

The idea of quark-lepton compositeness~\cite{Pati:1974yy,Terazawa:1976xx,Neeman:1979wp,Harari:1979gi,Shupe:1979fv,
Terazawa:1979pj,Harari:1980ez,Fritzsch:1981zh} goes along with our intention to describe nature in terms of its most fundamental building blocks. As its name suggests, in the models with quark-lepton compositeness, the Standard Model (SM) fermions are not  elementary but rather have finer substructures. Similarities between the SM lepton and quark sectors (like, both come with three flavors and behave similarly under the $SU(2)_{\rm L}\times U(1)_{\rm Y}$ gauge symmetry with the same weak coupling) can be explained if they are assumed to be different bound states of some fundamental constituents. These fundamental constituents, called preons by Pati and Salam~\cite{Pati:1974yy}, are charged under some new strong force which confines them below a certain scale $\Lm$, known as the compositeness scale. 

As we have hadrons in QCD, in this scenario one expects a host of new exited preonic-condensates. Some of these condensates would be quite exotic, as they would carry both $SU(3)_{\rm c}$ color charges and lepton numbers, like the bosonic leptoquarks (LQs or $\lq$'s) that transform as triplets under $SU(3)_{\rm c}$~\cite{Buchmuller:1986zs,Hewett:1997ce,Kramer:1997hh} or the leptogluons (LGs or $\lg$'s) that are color-octet fermions \cite{Harari:1985cr,Baur:1985ud,Nir:1985ah,Rizzo:1985dn,Rizzo:1985ud,Streng:1986my} etc. 
Because of their color charges, if these exotic condensates have TeV-range masses, they would be produced copiously at the Large Hadron Collider (LHC)  making it possible to probe this scenario experimentally. 

The LHC has already put some constraints on the masses of scalar LQs decaying to SM quarks and leptons 
\cite{Aad:2015caa,Khachatryan:2015vaa,Khachatryan:2015bsa,Khachatryan:2015qda}.
Of these, we look at the most recent search by CMS, for the first and second generations of scalar LQs in the $\ell\ell jj$ and the $\ell\n_\ell jj$ channels with 19.7 fb$^{-1}$ of integrated luminosity at the 8 TeV LHC~\cite{Khachatryan:2015vaa}.  With pair production, the 95\% confidence level (CL) exclusion limit on the mass of the first (second) generation scalar LQ now stands at $M_{\lq} = 1005$ (1080) GeV assuming it always decays to an electron (a muon) and a jet. Note that unless specified otherwise, we do not distinguish between any particle and its anti-particle. Hence, an electron here could mean a positron as well. In the first generation search, mild excesses of events compared to the SM background  were observed in both the $eejj$ and the $eej$ channels for $M_{\lq}\sim$ 650 GeV. Currently, these excesses have attracted considerable attention in the  literature. CMS has also performed a dedicated search for the single productions of the first two generations of LQs in the $\ell\ell j$ channels~\cite{Khachatryan:2015qda}. However, unlike the mostly QCD mediated pair production, the single productions depend strongly on an unknown coupling $\lm$, the $\lq$-$\ell$-$q$ coupling. Hence, the exclusion limits from this search are $\lm$ dependent. For the first generation, the exclusion limit goes from 895 GeV to 1730 GeV when  $\lm$ goes from 0.4 to 1.0 and for the second generation the data exclude $M_{\lq}$ below 530 GeV for $\lm=1.0$. 

In this letter, we recast the CMS 8 TeV $eejj$~\cite{Khachatryan:2015vaa} and $eej$~\cite{Khachatryan:2015qda} data in terms of the first generation spin-1/2 LG carrying unit electric charge, {\it i.e.}, the color octet partner of the SM electron ($e_8$) to probe the composite quark-lepton scenarios and obtain the most stringent limits available on the $e_8$. This is possible because a LG can also decay to a lepton and a jet (gluon) just like a LQ. Hence, the pair production of $e_8$'s would have $eejj$ final states.\footnote{In absence of any BSM decay, the only two body decay a LG can have is either $\lg\to\ell\ g$ or $\n_8\to\n_\ell\ g$ ($\n_8$, color octet partner of a neutrino) but not both. Hence, unlike LQs, the QCD mediated pair production of LGs can not have a $\ell\n_{\ell}jj$ final state. However, depending on the underlying model, a charged $\ell_8$ and a neutral $\n_8$ might couple simultaneously with a SM $W$ boson allowing a weak interaction mediated process,
\ba
pp\to (W\to \ell_8\n_8) \to \ell j\ubr{-2.5}\ \n_{\ell} j\lbr{-2.5}\ ,\nn
\ea
with the $\ell\n_{\ell} jj$ final state. 
} Earlier, there have been other phenomenological studies on LGs~\cite{Celikel:1998dj,Sahin:2010dd,Akay:2010sw,Jelinski:2015epa,Acar:2015wxp} and the CMS 7 TeV $eejj$ data~\cite{Chatrchyan:2012vza} were used to infer bounds on $M_{e_8}$~\cite{Goncalves-Netto:2013nla,Mandal:2012rx}. Considering the pair production, Ref.~\cite{Goncalves-Netto:2013nla} put the mass exclusion limit at about 1.2-1.3 TeV.  Similarly, an $e_8$ could be produced singly in association with an electron and give rise to an $eej$ final state. Interestingly, the single productions of LGs 
open up a way to probe the compositeness scale. This is because, at the leading order (LO), the $\lg$-$\ell$-$g$ interaction comes from an effective operator of dimension five that is suppressed by the compositeness scale $\Lm$~\cite{Agashe:2014kda,Mandal:2012rx} (see the next section). This is unlike the LQ interactions, where the LO terms are of dimension four and hence, apparently insensitive to $\Lm$.

In a recent paper~\cite{Mandal:2015vfa}, we pointed out that the single productions of LQs can also lead to the $eejj$ final state and similarly, events from the pair productions could also pass the signal selection criteria of the single production search in the $eej$ channel. Combining these production processes in the signal simulations can provide better limits in the $M_{\lq}$-$\lm$ plane from both the $eejj$ and the $eej$ channels. The same argument applies for LGs too. Hence, following Ref.~\cite{Mandal:2015vfa}, here we systematically combine both the pair and the single production processes of the  $e_8$ while reinterpreting the CMS $eejj$ and $eej$ data and obtain exclusion limits in the $M_{e_8}$-$\Lm$ plane. This way, we obtain the mass exclusion limits as well as the limits on the compositeness scale from both the $eejj$ and the $eej$ data and compare them.

Our presentation is organized as follows. In the next section we discuss the details of the signal we consider, in section \ref{sec:three}, we present the results of our recast analysis, in section \ref{sec:futpros} we investigate the prospect of discovering the color octet electron at the 13 TeV LHC and then in section \ref{sec:last} we conclude.

\section{Leptogluon (Combined) Signals}\label{sec:two}

\begin{figure*}[!t]
\subfloat[]{\includegraphics[scale=0.45]{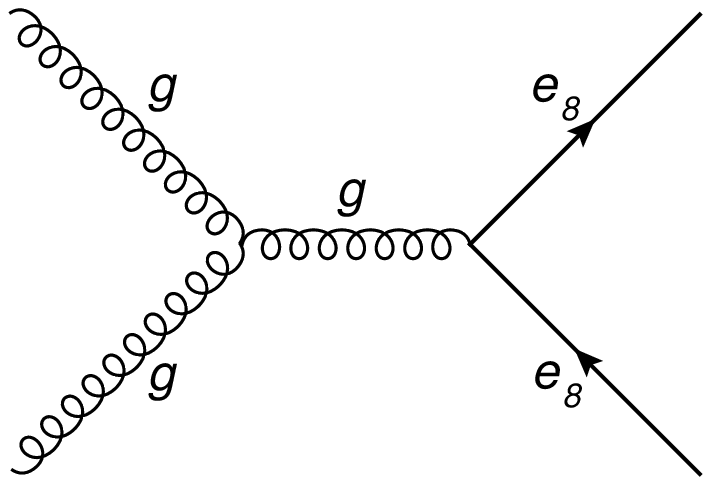}\label{fig:feyn_pair_1}}\hfill
\subfloat[]{\includegraphics[scale=0.45]{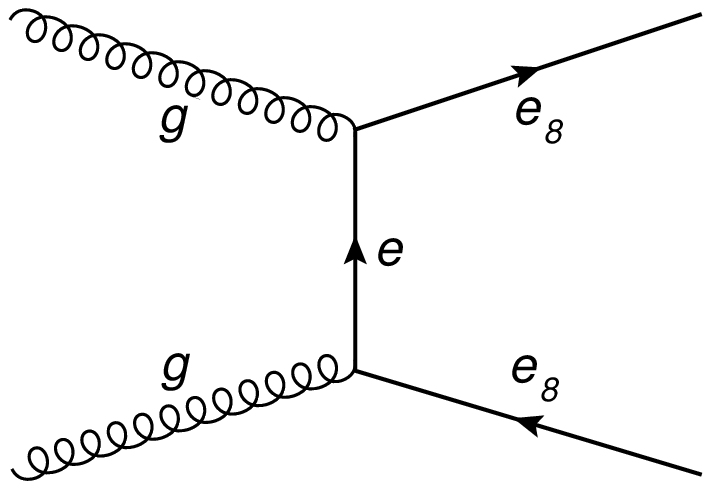}\label{fig:feyn_pair_e_ex}}\hfill
\subfloat[]{\includegraphics[scale=0.45]{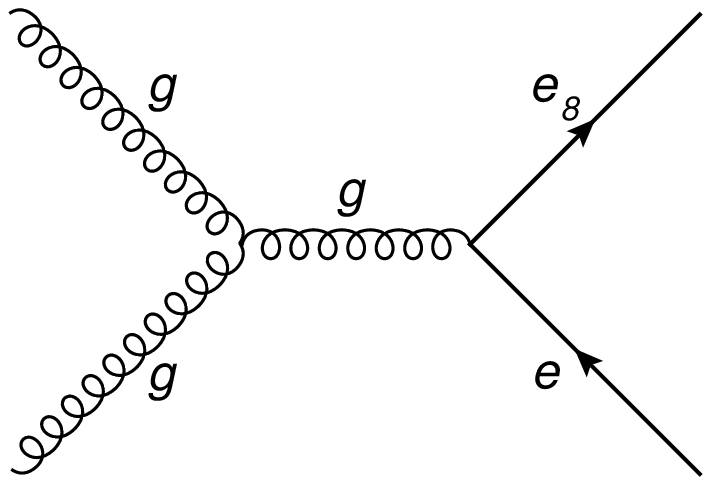}\label{fig:feyn_sin_2bd}}\hfill
\subfloat[]{\includegraphics[scale=0.45]{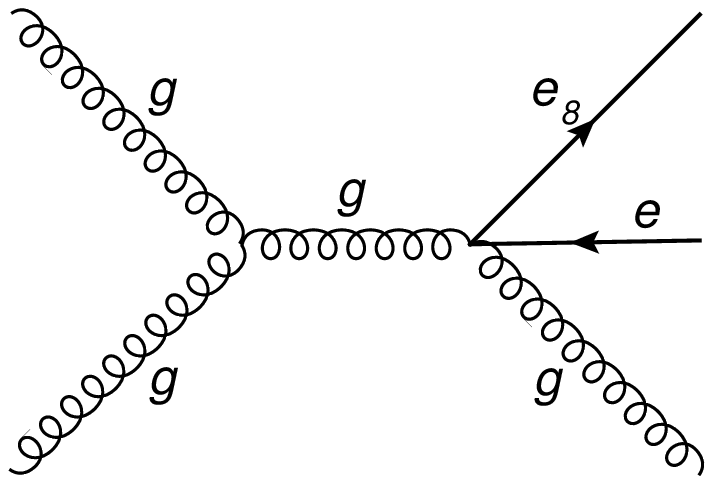}\label{fig:feyn_sin_3bd}}
\caption{Sample Feynman diagrams: (a) \& (b) LG pair production, (c) \& (d) LG single productions at the LHC.
}
\label{fig:fyndiags}
\end{figure*}

If we assume $M_{e_8}$ is smaller than $\Lm$ and there is no violation of lepton flavor, we can write a generic effective Lagrangian for the $e_8$ allowed by the SM 
gauge symmetry as~\cite{Mandal:2012rx},
\ba
\mc{L} = \bar e_8^a i\gm^\mu\left(\partial_\mu\dl^{ac} + g_s f^{abc}G^b_\mu \right) e_8^c 
- M_{e_8}\bar{e}_8^a e_8^a + \mc{L}_{\rm int},\label{eq:lag}
\ea
with~\cite{Agashe:2014kda},
\ba
\mc{L}_{\rm int} = \frac{g_s}{2\Lm} G^a_{\m\n} \left[\bar{e}_8^a \sg^{\mu\nu}\left(\eta_L e_{L} + \eta_R e_{R}\rt)\rt] + \mbox{H.c.} + \,\ldots.\label{eq:intlag}
\ea
In the Lagrangian, we have displayed only those dimension five terms that are important for our study.\footnote{As pointed out in Ref.~\cite{Mandal:2012rx}, there are more dimension five operators allowed by the SM gauge
symmetries and lepton number conservation like,
\ba
\frac{\mc{C}_8}{\Lm} i f^{abc} \bar{e}_8^a G_{\mu\nu}^{b} \sg^{\mu\nu} e_8^c +  
\frac{\mc{C}_1}{\Lm} \bar{e}_8^a B_{\mu\nu} \sg^{\mu\nu} e_8^a\ .\nn
\ea
However, these terms lead to $e_8e_8V$ or $e_8e_8VV$ vertices (may contain form factors) that would affect the production cross section. 
For simplicity, we assume the unknown coefficients associated 
with these terms are negligible.\label{fn:footnote1}}
Here, $G^a_{\mu\nu}$ is the gluon field strength tensor,  and $\et_{L/R}$ are the chirality factors. 
Since, the electron chirality conservation implies $\et_L\et_R = 0$, we set 
$\et_L = 1$ and $\et_R=0$ in our analysis without any loss of generality. This dimension five interaction opens two decay modes for the color octet electron: $e_8\to eg$ and $e_8\to egg$. However, since the three body decay is more suppressed than the two body one, we simply set the total width of the $e_8$ as~\cite{Mandal:2012rx,Celikel:1998dj},
\ba
\Gm \approx \frac{\al_s\lt(M_{e_8}\rt)M_{e_8}^3}{4\Lm^2}. 
\ea

The production processes of the $e_8$ at the LHC (see Fig.~\ref{fig:fyndiags} for some representative Feynman diagrams) are discussed in much detail in Ref.~\cite{Mandal:2012rx}. Instead, here we focus on some essential points. The main contribution to the $e_8$ pair production comes from the purely QCD mediated diagrams (see {\em e.g.} Fig.~\ref{fig:feyn_pair_1}). At the LO, there is an additional $t$-channel electron exchange diagram whose amplitude is proportional to $1/\Lm^2$ (Fig.~\ref{fig:feyn_pair_e_ex}) but, for the ranges of $M_{e_8}$ and $\Lm$ we consider in this letter, its contribution is small compared to the model independent QCD mediated contribution. That is why the pair production process is practically insensitive to the compositeness scale. On the other hand,  all the single production diagrams contain at least one $e_8$-$e$-$g$ or $e_8$-$e$-$g$-$g$ vertex ($\sim 1/\Lm$) coming from the interaction term of Eq. \eqref{eq:intlag} (see {\em e.g.} Figs.~\ref{fig:feyn_sin_2bd} \&~\ref{fig:feyn_sin_3bd}).

We simulate the pair and the single productions of $e_8$ at the 8 TeV LHC to estimate their contributions to the $eejj$ and the $eej$  channels by modeling Eqs.~\eqref{eq:lag} and \eqref{eq:intlag} in \textsc{Feynrules}~\cite{Alloul:2013bka}. We use the CTEQ6L1 Parton Distribution Functions (PDFs)~\cite{Pumplin:2002vw} to generate events with \textsc{MadGraph5} \cite{Alwall:2014hca} and then shower them with \textsc{Pythia}6 \cite{Sjostrand:2006za}. 
We set the factorization and the renormalization scales, $\mu_{\rm F}$ $=\mu_{\rm R}$ $=M_{e_8}$. We use {\sc Delphes 3.3.1} \cite{deFavereau:2013fsa} to simulate the CMS detector environment and implement the selection cuts. In {\sc Delphes}, jets are clustered with {\sc FastJet}~\cite{Cacciari:2011ma} using the anti-$k_{\rm T}$ jet
clustering algorithm~\cite{Cacciari:2008gp} with the clustering parameter, $R=0.4$. Since, we generate the pair and the single productions separately, any possible interference between them has been ignored. However, this is justified as, for the parameters considered, the $e_8$ decay width is much smaller than its mass ({\it i.e.}, narrow width regime).

We generate events for the inclusive single production for certain $\Lm=\Lm_{\rm o}$ by combining the following processes,
\be
\left[ \begin{array}{lclcl}
pp &\to &(e_8\ e) 		&\to & e j\ubr{-2.5}\ e\,,\\
pp &\to &(e_8\ ej)		&\to & e j\ubr{-2.5}\ ej\,,\\
pp &\to &(e_8\ ejj)  	&\to & e j\ubr{-2.5}\ ejj\,,
\end{array}\right]_{\Lm=\Lm_{\rm o}}\label{eq:matching_ee}
\ee
where the curved connections indicate a pair of electron and gluon coming from an on-shell $e_8$. However, a straightforward computation of cross section for the combined single and pair production processes would lead to some difficulties. Like, the jets that are not coming from a LG could be soft and lead to divergences. Ideally, to handle these divergences, one has to go beyond a tree level computation while combining the different single production processes as in Eq.~\eqref{eq:matching_ee}. Moreover, such combination can lead to double counting of some diagrams while showering. Following Ref.~\cite{Mandal:2015vfa}, we avoid these difficulties by employing the matrix element-parton shower matching (ME$\oplus$PS) technique  
with the shower-$k_{\rm T}$ scheme~\cite{Alwall:2007fs,Alwall:2008qv} 
which effectively provides a consistent interpolation between the hard partons and the {\sc Pythia} parton showers (PS). It relies on the {\sc Pythia} PS for the soft jets and the parton level matrix elements for the hard jets and thereby, bypasses the double counting and the soft jets problems. 

The cross section for any other value of $\Lm=\Lm_{\rm n}$ (say) is obtained by simply multiplying the cross section for $\Lm_{\rm o}$ by $\Lm_{\rm o}^2/\Lm_{\rm n}^2$, since, as explained earlier, the  $\Lm$ dependence of the inclusive single production cross section ($\sg^s$) can be written as,
\ba
\sg_{\rm s}\left(M_{e_8},\Lm\right) \overset{\rm def.}{=} \frac{1}{\Lm^2}\bar\sg_s\left(M_{e_8}\right),\label{eq:sigmabar}
\ea
if we ignore terms of  $\mc O\left(1/\Lm^4\right)$ or higher.
In Table \ref{tab:xsec}, we show  $\sg_{\rm s}( M_{e_8},\Lm)$ for four difference choices of $M_{e_8}=$ 0.5, 1.0, 1.5 \& 2.0 TeV and two different choices of $\Lm=$ 2.5 \& 5 TeV.  There, we also show the LO values of the pair production cross-section ($\sg_{\rm p}^{\rm LO}$) for the four masses. While combining the pair and the single productions, we use the next-to-leading (NLO) in QCD $K$-factors only for the pair production,  available from Ref.~\cite{Goncalves-Netto:2013nla} for masses up to 1.5 TeV. Beyond this, guided by the trend, we assume a constant $K_{\rm NLO}=$ 2.0.~\footnote{As it is clear from Table~\ref{tab:xsec}, $\sg_{\rm p}$ is too small for $M_{e_8} \gtrsim$ 1.5. Hence, in practice, this assumption matters little, though we make it for consistency.}
Note, however, no $K$-factor is available for the single productions. Hence, for a particular $\Lm$, we utilize the available information to the best possible manner and use the combined signal with the following cross section,
\ba
\hspace{-.5cm}&&\hspace{-1.cm} \sg_{\rm p}\left(M_{e_8}\right) + \frac{1}{\Lm^2}\bar\sg_{\rm s}\left(M_{e_8}\right)\nn\\
&&\hspace{-.5cm} =\ K_{\rm NLO}\left(M_{e_8}\right)\times \sg_{\rm p}^{\rm LO}\left(M_{e_8}\right)  + \sg_{\rm s}\left(M_{e_8},\Lm\right).\label{eq:comxsec}
\ea

\begin{table}[!t]
\begin{tabular}{|>{\centering}m{0.24\linewidth} |>{\centering}m{0.24\linewidth} |>{\centering}m{0.24\linewidth} c|}
\hline 
$M_{e_8}$  & $\sg_{\rm p}^{\rm LO}$ (fb) &  \multicolumn{2}{c|}{$\sg_{\rm s}$ (fb)}\\
\cline{3-4}
(TeV)  & ($\Lm\to\infty$) & $\Lm=$ 2.5 TeV & $\Lm=$ 5.0 TeV \\ 
\hline \hline
0.5  & 3.85$\times 10^{3}$  & 4.85$\times 10^{3}$ & 1.22$\times 10^{3}$ \\ 
1.0  & 1.77$\times 10^{1}$ & 2.66$\times 10^{1}$ & 6.60$\times 10^{0}$ \\ 
1.5  & ~~2.36$\times 10^{-1}$ & 3.03$\times 10^{0}$ & ~~7.62$\times 10^{-1}$ \\
2.0  & ~~3.22$\times 10^{-3}$ & ~~4.41$\times 10^{-1}$ & ~~1.09$\times 10^{-1}$\\
\hline 
\end{tabular} 
\caption{Cross sections of the pair ($\sg_{\rm p}$, at the LO) and the inclusive single ($\sg_{\rm s}$, generated as shown in Eq.~\eqref{eq:matching_ee}) productions of color octet electrons at the 8 TeV LHC. }\label{tab:xsec}
\end{table}

\begin{figure*}[!t]
\subfloat[]{\includegraphics[scale=0.610]{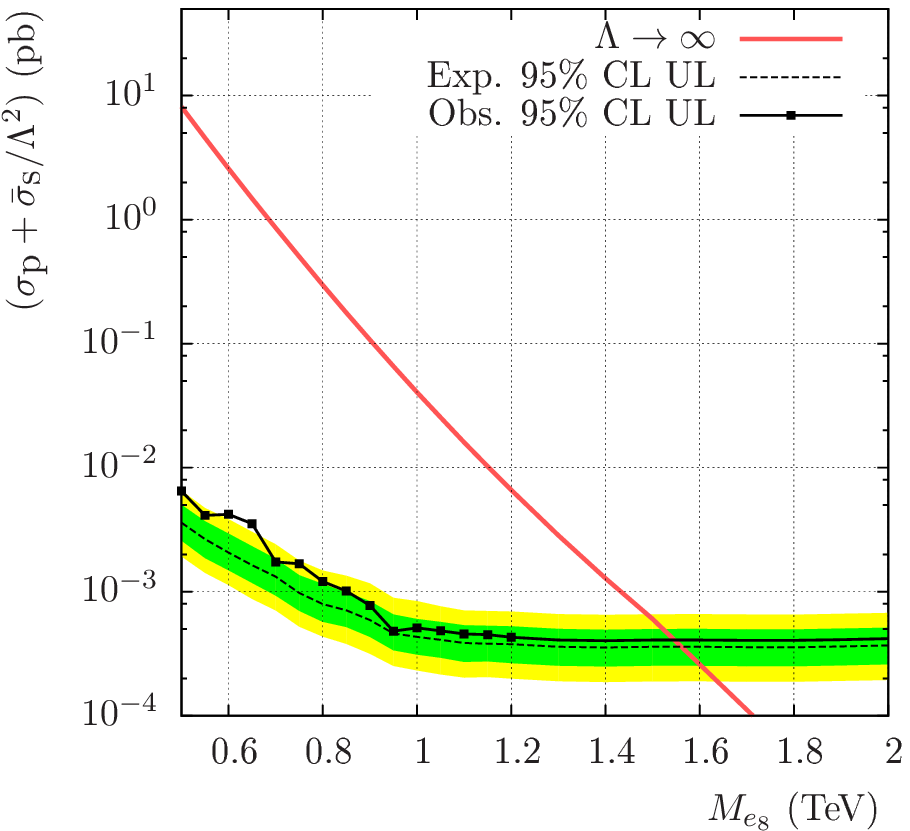}\label{fig:eejj_data_a}}\hfill
\subfloat[]{\includegraphics[scale=0.610]{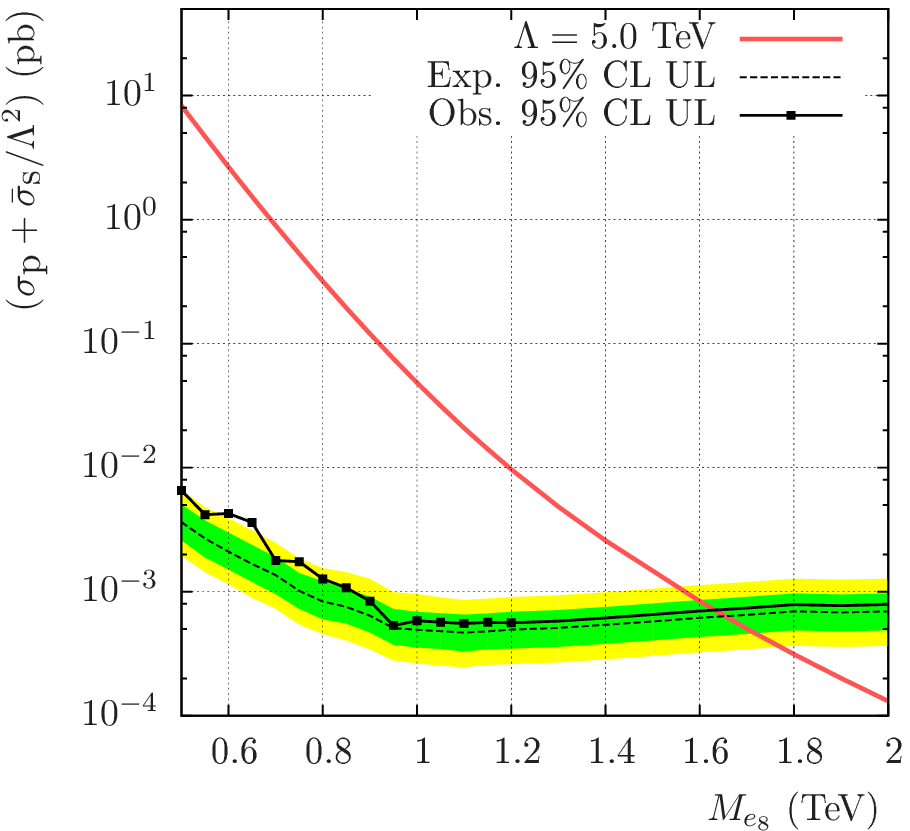}\label{fig:eejj_data_b}}\hfill
\subfloat[]{\includegraphics[scale=0.610]{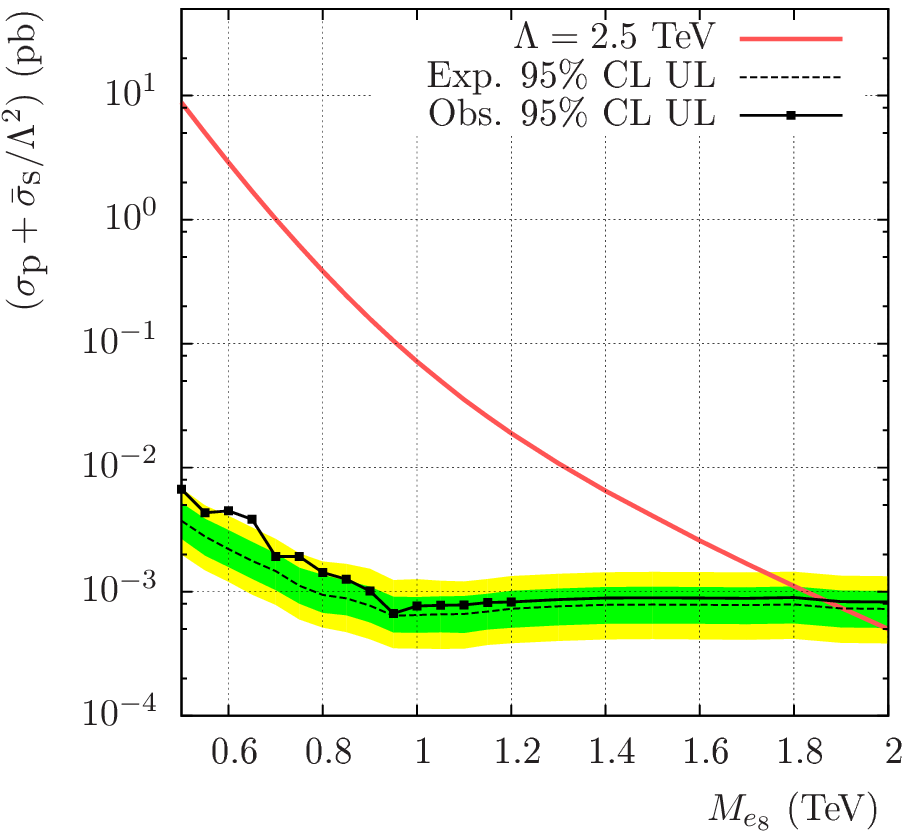}\label{fig:eejj_data_c}}
\caption{Mass exclusion plots from the recast analysis of the $eejj$ data~\cite{Khachatryan:2015vaa} with three different choices of the compositeness scale $\Lm$. For these plots, combined production of color octet electrons ({\it i.e.}, the QCD mediated pair production plus the inclusive single productions from  Eqs.~\eqref{eq:matching_ee} \&~\eqref{eq:sigmabar}) with cross section (solid red lines) $\sg_{\rm p} + \bar\sg_{\rm s}/\Lm^2$, Eq.~\eqref{eq:comxsec}, is considered to simulate the signal. 
To obtain the expected 95\% CL upper limits (dashed lines) beyond 1.2 TeV, the selection cuts~\cite{Khachatryan:2015vaa} are assumed to be identical for $M_{e_8} \geq 1.2$ TeV.
}
\label{fig:eejj_data}
\end{figure*}

\begin{figure*}[!t]
\subfloat[]{\includegraphics[scale=0.610]{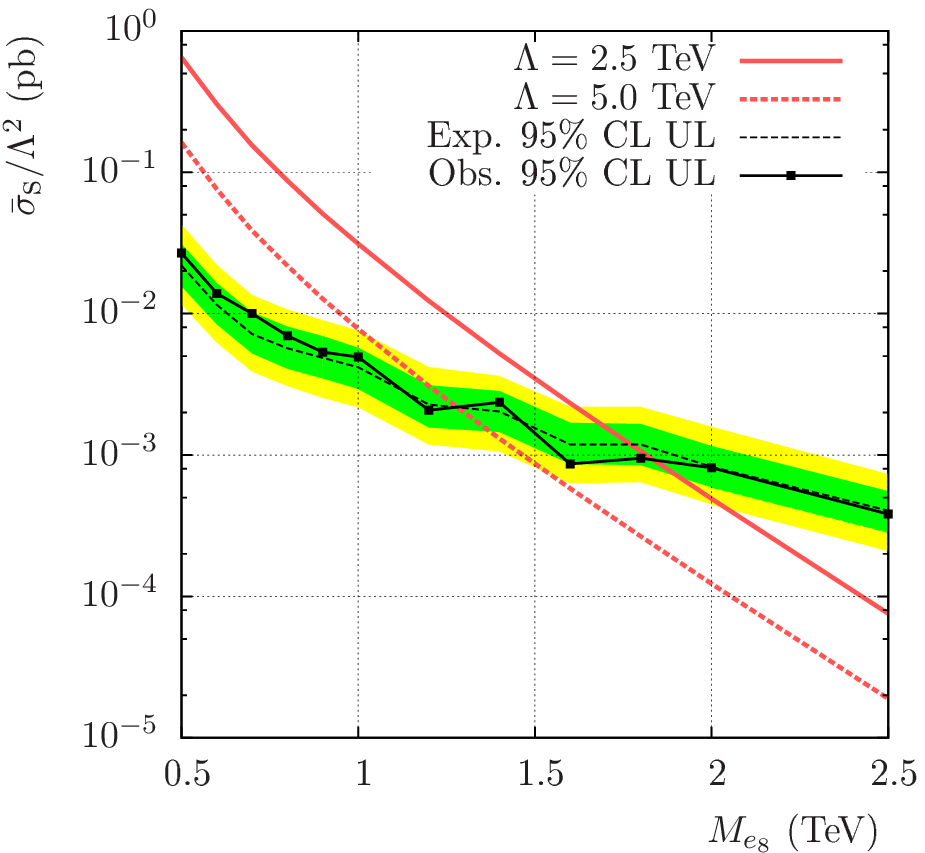}\label{fig:eej_data_a}}\hfill
\subfloat[]{\includegraphics[scale=0.610]{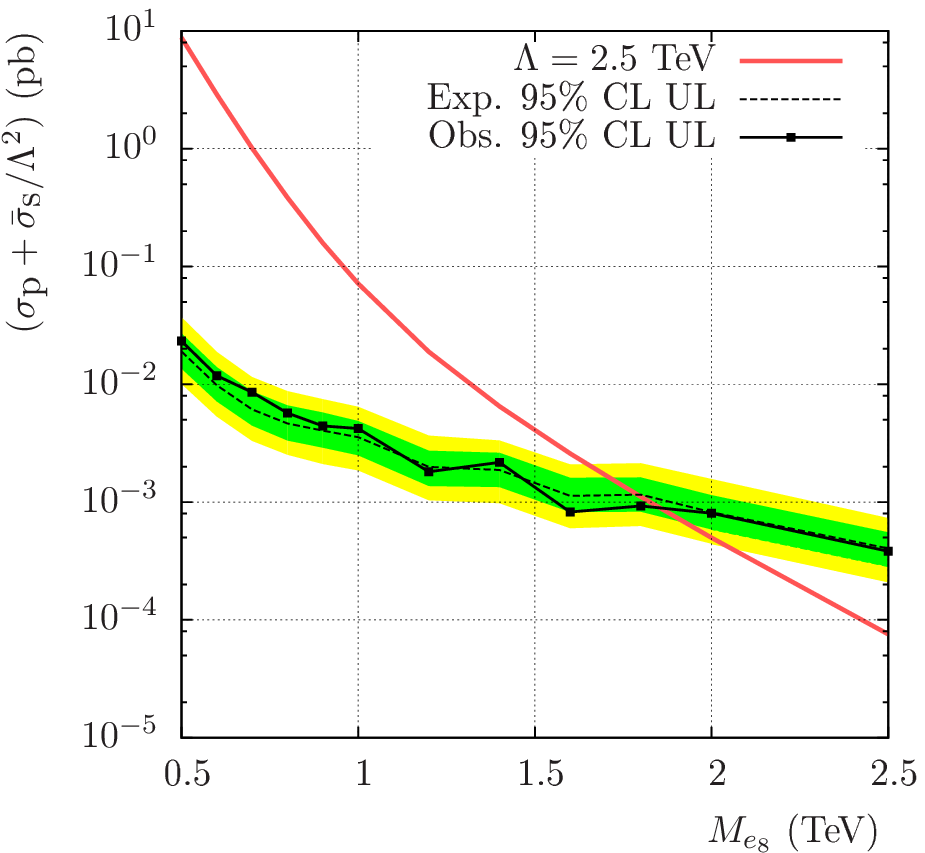}\label{fig:eej_data_b}}\hfill
\subfloat[]{\includegraphics[scale=0.610]{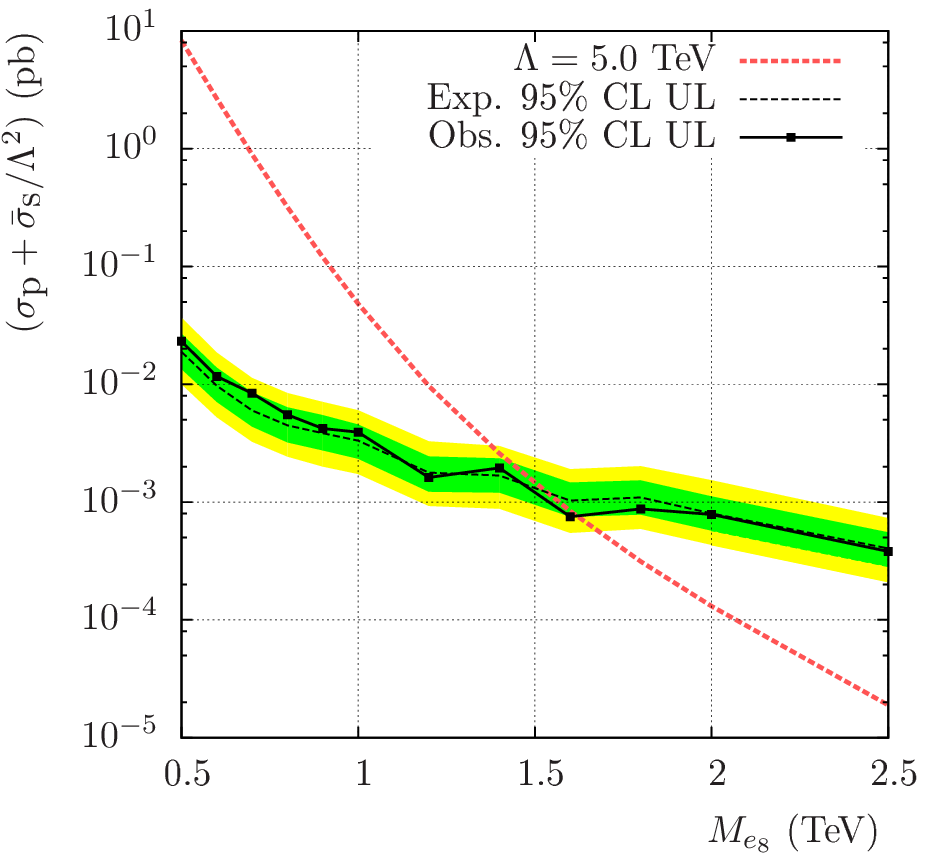}\label{fig:eej_data_c}}
\caption{Mass exclusion plots from the recast analysis of the $eej$ data~\cite{Khachatryan:2015vaa} for two different $\Lm$'s. In (a), only $e_8$ single production is considered in the signal while in (b) and (c), single and pair productions are combined following Eq.~\eqref{eq:comxsec} to simulate the signal. 
}
\label{fig:eej_data}
\end{figure*}

\begin{figure*}[!t]
\subfloat[]{\includegraphics[scale=0.610]{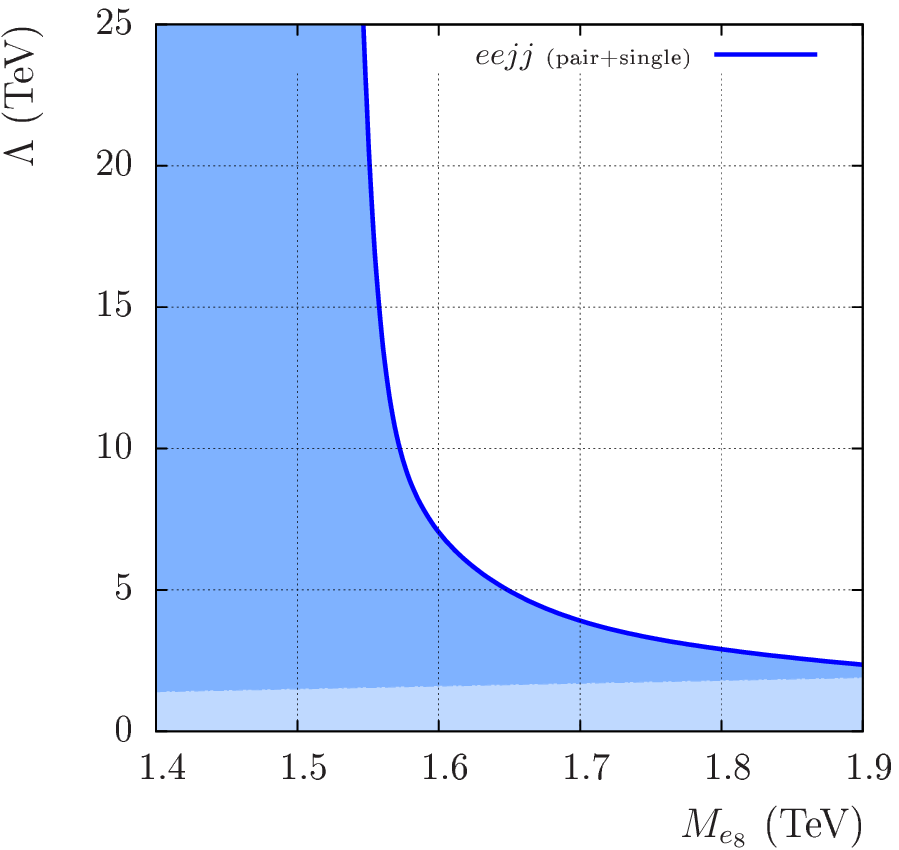}\label{fig:lm-me8-limit_a}}\hfill
\subfloat[]{\includegraphics[scale=0.610]{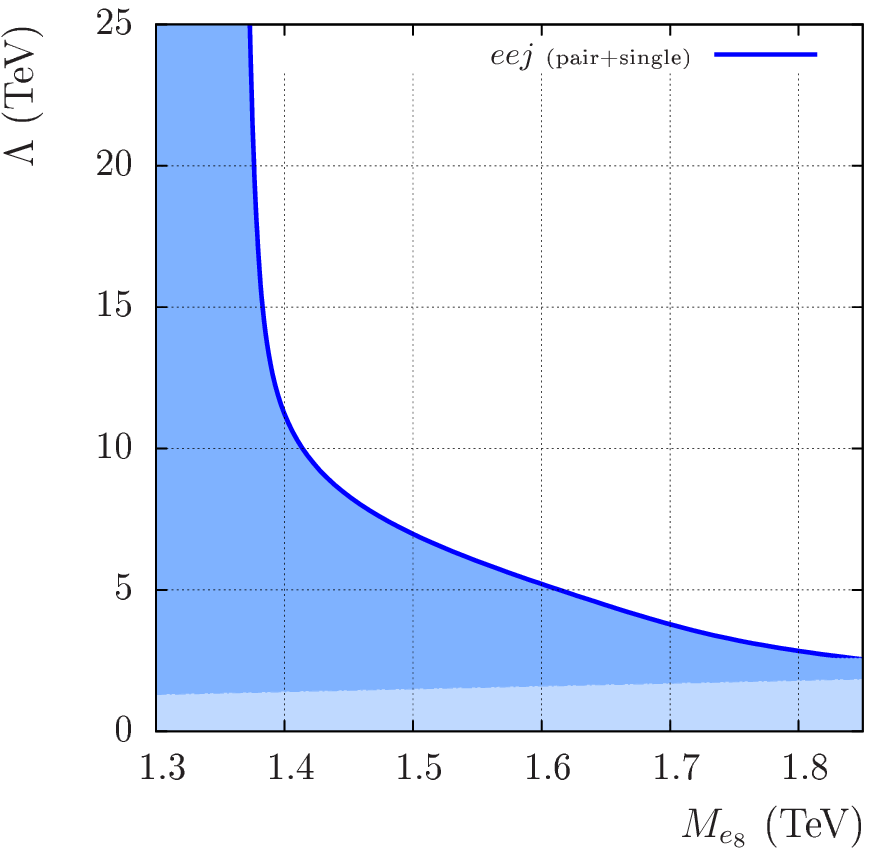}\label{fig:lm-me8-limit_b}}\hfill
\subfloat[]{\includegraphics[scale=0.610]{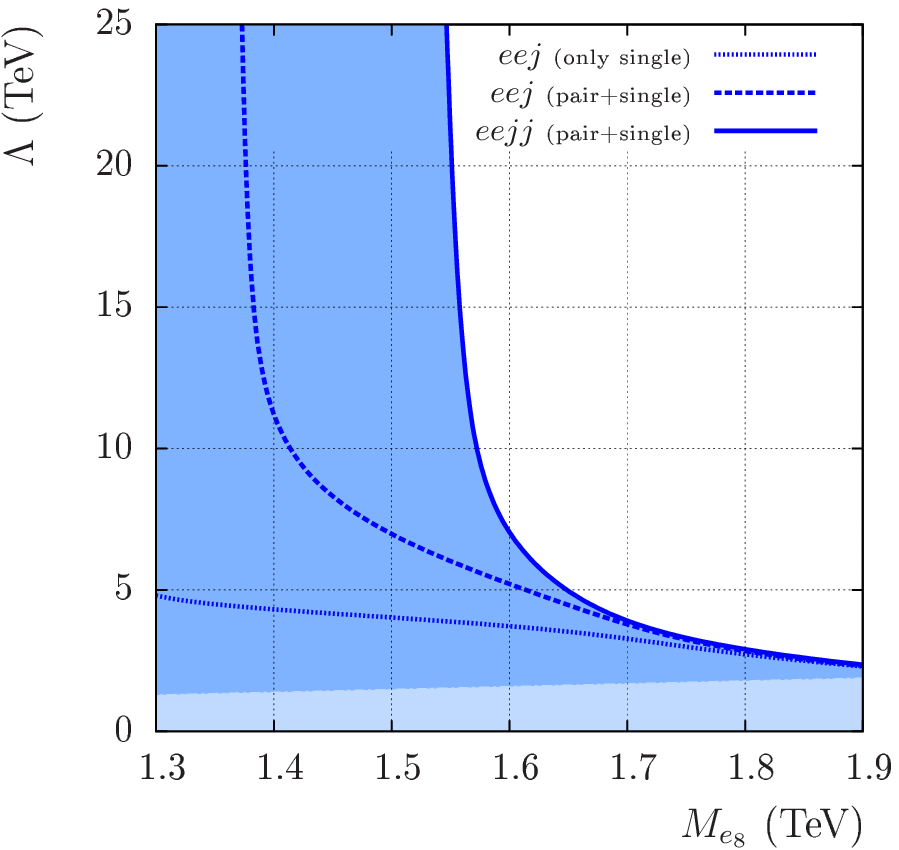}\label{fig:lm-me8-limit_c}}
\caption{Exclusion limits in the $M_{e_8}$-$\Lm$ plane: (a) from the CMS $eejj$ data ~\cite{Khachatryan:2015vaa} (obtained with combined production), (b) from the CMS $eej$ data~\cite{Khachatryan:2015vaa} (obtained with combined production) and (c) comparison plot. The dark shaded regions are ruled out by the data. The  lightly shaded regions correspond to $M_{e_8}>\Lm$ where our effective theory approach (Eqs.~\eqref{eq:lag} \& \eqref{eq:intlag}) is not reliable.}
\label{fig:lm-me8-limit}
\end{figure*}

\section{Recast Analysis and New Limits}\label{sec:three}

In Fig.~\ref{fig:eejj_data}, we show the recast mass exclusion plots obtained from the CMS $eejj$~\cite{Khachatryan:2015vaa} data for three different values of $\Lm$, namely, $\Lm\to\infty$ ({\it i.e.} the pair production only) in Fig.~\ref{fig:eejj_data_a}, $\Lm=5$ TeV in Fig.~\ref{fig:eejj_data_b} and $\Lm=2.5$ TeV in Fig.~\ref{fig:eejj_data_c}. To obtain the expected and the observed 95\% CL upper limits (ULs) for the recast plot,  we rescale the corresponding limits from the CMS plot~\cite{Khachatryan:2015vaa} by multiplying with a factor~\cite{Mandal:2015vfa},
\ba
\mc R^{eejj}_{\lq\to e_8}\left(M_{e_8},\Lm\right) =  \frac{\ep_{\rm p}^{(\lq|eejj)}\lt(M_{\lq}=M_{e_8}\rt)}{\ep_{\rm p+s}^{(e_8|eejj)}\lt(M_{e_8},\Lm\rt)}\,,\label{eq:efficiency_rescale_eejj}
\ea
where $\ep_{\rm p}^{(\lq|eejj)}\lt(M_{e_8}\rt)$  is the efficiency (yield) of the final event selection cuts optimized for the pair production of the  first generation scalar LQ of mass $M_{\lq}=M_{e_8}$~\cite{Khachatryan:2015vaa} and $\ep_{\rm p+s}^{(e_8|eejj)}\lt(M_{e_8},\Lm\rt)$ is the efficiency of the same set of cuts estimated for the combined (pair+single, combined as in Eq.~\eqref{eq:comxsec}) productions of $e_8$'s. In other words, $\ep_{\rm p+s}^{(e_8|eejj)}\lt(M_{e_8},\Lm\rt)$ denotes the fraction of the combined signal events that survives the selection cuts optimized for $M_{\lq}=M_{e_8}$. Since, the CMS $eejj$ optimized cuts stop at $M_{\lq}=1.2$ TeV, we extrapolate beyond this mass by assuming identical selection cuts for $M_{\lq}\geq 1.2$ TeV. Because of the single productions, the lower limit of the allowed mass increases with decreasing $\Lm$. For example, from the  pure QCD mediated pair production ($\Lm\to\infty$) the limit stands at about 1.56 TeV and it improves to about 1.66 (1.90) TeV for $\Lm=5~(2.5)$ TeV.\footnote{Production processes for LGs generally have enhanced color factors than LQ production processes (color octet LGs vs. color triplet LQs). As a result, from the same data one generally obtains higher mass exclusion limits for LGs than LQs for similar choice of parameters.} Note that with increasing mass, the pair production becomes more phase space suppressed compared to the single productions and hence, beyond a certain mass, the single productions dominate over the pair production. The crossover point depends on $\Lm$, since all the single productions depend on it. With this in mind, we can now understand the behavior shown by the 95\% CL UL lines in the high $M_{e_8}$ limit for finite $\Lm$'s. We expect the single productions to take over the pair production earlier when $\Lm=2.5$ TeV than when $\Lm=5$ TeV. This can be seen from Figs.~\ref{fig:eejj_data_b} \& \ref{fig:eejj_data_c}: the small raise in any UL line with increasing $M_{e_8}$ (that it is indeed coming from the single productions  can be confirmed from its absence in the pair only plot) comes earlier for $\Lm=2.5$ TeV than $\Lm=5$ TeV.~\footnote{It is not very straight forward to understand the reason behind  the raise itself intuitively. When these selection cuts~\cite{Khachatryan:2015vaa} are held fixed, both the efficiencies  start to increase with increasing $M_{e_8}$ till they saturate. However, since they evolve differently, there is a competition between the numerator and the denominator of Eq.~\eqref{eq:efficiency_rescale_eejj}.}

In Fig.~\ref{fig:eej_data}, the recast plots for $\Lm=2.5$ and 5 TeV obtained from the CMS $eej$~\cite{Khachatryan:2015vaa} data are shown. For Fig.~\ref{fig:eej_data_a}, we have considered only the single productions in the signal to compare the mass exclusion limits for the two values of $\Lm$ while in Figs.~\ref{fig:eej_data_b} \&~\ref{fig:eej_data_c}, we consider the combined productions. Here, we rescale the CMS limits~\cite{Khachatryan:2015vaa} by
\ba
\tilde{\mc R}^{eej}_{\lq\to e_8}\left(M_{e_8}\right) =  \frac{\ep_{\rm s}^{(\lq|eej)}\lt(M_{\lq}=M_{e_8}\rt)}{\ep_{\rm s}^{(e_8|eej)}\lt(M_{e_8}\rt)}\label{eq:efficiency_rescale_eej_s}
\ea
for the single only plot (Fig.~\ref{fig:eej_data_a}) and by
\ba
\mc R^{eej}_{\lq\to e_8}\left(M_{e_8},\Lm\right) =  \frac{\ep_{\rm s}^{(\lq|eej)}\lt(M_{\lq}=M_{e_8}\rt)}{\ep_{\rm p+s}^{(e_8|eej)}\lt(M_{e_8},\Lm\rt)}\label{eq:efficiency_rescale_eej}
\ea
for the other two (Figs.~\ref{fig:eej_data_b} \&~\ref{fig:eej_data_c}). Here, $\ep_{\rm s}^{(\lq|eej)}\lt(M_{e_8}\rt)$  is the efficiency of the final event selection cuts optimized for the single productions of the  first generation scalar LQ of mass $M_{\lq}=M_{e_8}$~\cite{Khachatryan:2015vaa}.
Notice that though the single productions of the LQ depend on the unknown $\lq$-$\ell$-$q$ coupling $\lm$, the efficiency $\ep_{\rm s}^{(\lq|eej)}$, being a ratio of the number of events, does not depend on any overall factor in the cross section like $\lm$~\cite{Mandal:2015vfa}. For the same argument $\ep_{\rm s}^{(e_8|eej)}$, which is the cut efficiency for the inclusive single production of the $e_8$, does not depend on $\Lm$ even though $\ep_{\rm p+s}^{(e_8|eej)}$ does. If we compare Fig.~\ref{fig:eej_data_a} with Figs.~\ref{fig:eej_data_b} \&~\ref{fig:eej_data_c}, it is clear how the inclusion of the pair production in the signal for the $eej$ search improves the mass exclusion limits. For example, for $\Lm=5 (2.5)$ TeV the $eej$ data disfavor $M_{e_8}$ below 1.28 (1.84) TeV when only the single productions are considered. But the same limit goes up to about 1.62 (1.86) TeV when the pair production is also included. Obviously, the improvement is more prominent when the single productions are relatively smaller because of larger $\Lm$.

In Fig.~\ref{fig:lm-me8-limit}, we show the rescaled 95\% CL exclusion limits in the $M_{e_8}$-$\Lm$ plane. The blue shaded regions are disfavored by the data. We show the exclusion contours obtained from the CMS $eejj$ data (Fig.~\ref{fig:lm-me8-limit_a}) and the $eej$ data (Fig.~\ref{fig:lm-me8-limit_b}). We compare these two in Fig.~\ref{fig:lm-me8-limit_c}. The pair production dominates
in the lower mass region and gives a limit on $M_{e_8}$ that is practically independent of $\Lm$. From Fig.~\ref{fig:lm-me8-limit_a} or~\ref{fig:lm-me8-limit_c}, it is clear that irrespective of $\Lm$, the $eejj$ data 
disfavor the $e_8$ with mass below $\sim 1.55$ TeV. In the high mass region, the pair production becomes negligible and the inclusive single production puts a strong limit on $\Lm$. However, what is remarkable is that the $eejj$ data give almost identical limit as the $eej$ data in this regime. In other words, in the high mass limit, the contamination of single production in a search optimized for pair production is very significant.\footnote{Since the pair production search is insensitive to the spin of the particle being probed, kinematically it does not matter much whether the search is for LQs or LGs, at least in the narrow widths regime.} As explained in the introduction, the $\Lm$-dependent mass exclusion limits can also be translated as limits on $\Lm$. The overlapping limits in Fig.~\ref{fig:lm-me8-limit_c} indicate that the lightest limit on $\Lm$ stands about $\Lm\approx 2~\textrm{TeV}\approx M_{e_8}$ within the domain of the effective theory. If $M_{e_8}$ lies between 1.64 TeV and 2 TeV, $\Lm$ must be higher.

\section{Future prospects}
\label{sec:futpros}

So far our discussions were centered on reinterpreting the available data. Now, let us look at the prospect of a discovery of the $e_8$ at the LHC in its 13 TeV runs. In this section, we assume a future search in the $eej$ channel optimized for finding the $e_8$ and estimate the discovery reach using the combined production. 
We expect two high-$p_{\rm T}$
electrons and at least one high-$p_{\rm T}$ jet as the typical signature of the combined production
of $e_8$~\cite{Mandal:2012rx}. Therefore, taking a cue from the existing CMS $eej$ search~\cite{Khachatryan:2015vaa}, we use the following selection cuts:
\begin{enumerate}
\item two oppositely charged electrons ($e^{\pm}$) with transverse momentum $p_{\rm T}^e >$ 45 GeV 
and pseudorapidity $|\eta_e| <$ 2.1 excluding  1.442 $< |\eta_e| <$1.56,
\item the hardest jet must have $p_{\rm T}^{j_1} >$ 125 GeV \& $|\eta_{j_1}| < $ 2.4, 
\item separation between any electron and the hardest jet in the $\eta$-$\phi$ plane, $\Delta R_{ej_{1}} > $ 0.3.
\end{enumerate}
To suppress the inclusive-$Z$ background, we apply a strong cut on 
\begin{enumerate}\setcounter{enumi}{3}

\item the invariant mass of the electron pair, $M_{e_1e_2} >$ 400 GeV.
\end{enumerate}
In addition, we also apply some cuts optimized for the different $(M_{e_8},\Lm)$ combinations,
\begin{enumerate}\setcounter{enumi}{4}
\item
the scalar sum of the $p_{\rm T}$ of the two electrons and the hardest jet, 
\ba
S_{\rm T}=p_{\rm T}^{e_1}+p_{\rm T}^{e_2}+p_{\rm T}^{j_1} > S_{\rm T}^{\rm opt}\lt(M_{e_8},\Lm\rt),\label{eq:cut_st}
\ea
 
\item the maximum of the two electron-jet invariant mass combinations, 
\ba 
M_{ej}^{\rm max}={\rm Max\lt(M_{e_1j_1},M_{e_2j_1}\rt)}>M_{ej}^{\rm opt}\lt(M_{e_8},\Lm\rt).\label{eq:cut_mej}
\ea
\end{enumerate} 
The values of $S_{\rm T}^{\rm opt}$ and $M_{ej}^{\rm opt}$ for some benchmark parameters are shown in Table~\ref{tab:cut}.
The strong cut on $M_{e_1e_2}$ suppresses the inclusive $Z$ ($+n$ jets) contribution which is the most dominant background. 
The other significant
backgrounds are the inclusive top-pair production, the inclusive diboson ($ZZ$, $ZW$, $WW$) productions etc.~\cite{Mandal:2012rx}.

\begin{table}[!t]
\begin{tabular}{|>{\centering}m{0.1\linewidth}|>{\centering}m{0.135\linewidth}>{\centering}m{0.135\linewidth}|>{\centering}m{0.135\linewidth}>{\centering}m{0.135\linewidth}|>{\centering}m{0.135\linewidth} c|}
\hline
$M_{e_8}$   &  \multicolumn{2}{c|}{$\Lm\to\infty$} & \multicolumn{2}{c|}{$\Lm=$ 10 } & \multicolumn{2}{c|}{$\Lm=$ 5} \\
\cline{2-7}
  & $S_{\rm T}^{\rm opt}$ & $M_{ej}^{\rm max}$ & $S_{\rm T}^{\rm opt}$ & $M_{ej}^{\rm max}$ & $S_{\rm T}^{\rm opt}$ & $M_{ej}^{\rm max}$ \\
\hline \hline
1.5 &  0.7 &  0.5 &  0.7 &  0.5  & 0.5  &  0.5\\ 
2.0 & 1.4 &  0.6& 1.2 & 1.0  & 1.1 &  0.8 \\ 
2.5 & 1.8 & 1.8 & 1.7 & 1.7  & 1.5 & 1.2\\ 
3.0 & 1.8 & 1.8  & 1.8 & 1.8 & 1.8 & 1.8\\ 
\hline 
\end{tabular} 
\caption{Optimized $S_{\rm T}$ and $M_{ej}^{\rm max}$ cuts for the 13 TeV LHC. See Eqs.~\eqref{eq:cut_st} \& \eqref{eq:cut_mej} for the definitions of $S_{\rm T}^{\rm opt}$ and $M_{ej}^{\rm opt}$. All values are expressed in TeV.}
\label{tab:cut}
\end{table}
\begin{table}
\begin{tabular}{|>{\centering}m{0.1\linewidth}|>{\centering}m{0.135\linewidth}>{\centering}m{0.135\linewidth}|>{\centering}m{0.135\linewidth}>{\centering}m{0.135\linewidth}|>{\centering}m{0.135\linewidth} c|}
\hline
$M_{e_8}$   &  \multicolumn{2}{c|}{$\Lm\to\infty$} & \multicolumn{2}{c|}{$\Lm=$10 TeV} & \multicolumn{2}{c|}{$\Lm=$ 5 TeV} \\
\cline{2-7}
 (TeV) & Sig.& Backg.& Sig.& Backg. & Sig.& Backg. \\
 &   (fb)&(fb)&(fb)&(fb)&(fb)&(fb)\\
\hline \hline
1.5 & 9.385 & 2.764 & 10.253 & 2.764 & 13.035 & 3.786 \\ 
2.0 & 0.569 & 0.222 &  ~~0.771 & 0.282 &  ~~1.435 & 0.517 \\ 
2.5 & 0.039 & 0.025 &  ~~0.086 & 0.034 &  ~~0.263 & 0.105\\ 
3.0 & 0.003 & 0.025 &  ~~0.018 & 0.025 &  ~~0.065 & 0.025\\ 
\hline 
\end{tabular} 
\caption{Effects of optimized $S_{\rm T}$ and $M_{ej}^{\rm max}$ cuts on the combined signal and the dominant $Z+nj$ background (includes  contributions from $ZV$). The numbers show the cross sections computed for the 13 TeV LHC after applying the selection cuts.}
\label{tab:cut_eff}
\end{table}

\begin{figure}[!t]
\includegraphics[scale=0.75]{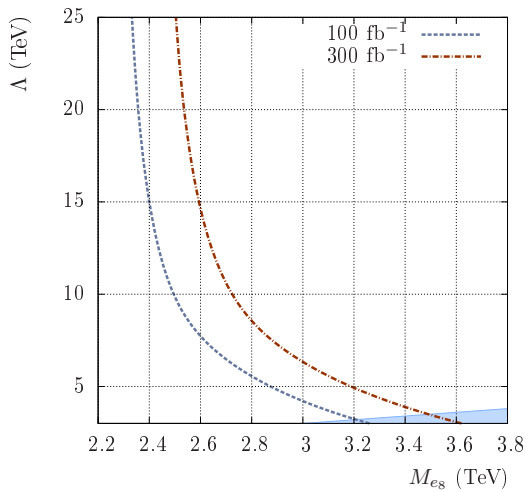}
\caption{The estimated 100 \& 300 fb$^{-1}$ contours of the discovery luminosity $\mc L_{\rm D}$ (Eq.~\eqref{eq:ld}) for the $e_8$ (first generation spin-1/2 leptogluon) at the 13 TeV LHC. The shaded region corresponds to $M_{e_8}>\Lm$ where our effective theory approach (Eqs.~\eqref{eq:lag} \& \eqref{eq:intlag}) is not reliable.}
\label{fig:LD}
\end{figure}

\begin{figure}[!t]
\includegraphics[scale=0.75]{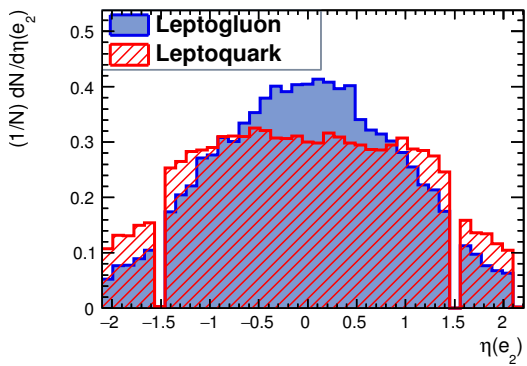}
\caption{The $\et$ distribution of the second hardest electron can be used to distinguish spin-0 LQs from spin-1/2 LGs. Here we set $M_{\lq}=M_{e_8}=$ 2 TeV and $\lm=$ 0.3 (for the LQ signal) and $\Lm=$ 10 TeV (for the LG signal) at the 13 TeV LHC.}
\label{fig:rapdist}
\end{figure}

To figure out the optimized values of the $S_{\rm T}$ and $M_{ej}$ cuts, {\it i.e.}, $S_{\rm T}^{\rm opt}$ and $M_{ej}^{\rm opt}$, we scan a square grid in the  
$S_{\rm T}$-$M_{ej}^{\rm max}$ plane  defined 
between 0.5 TeV \& 1.8 TeV in steps of 0.1 TeV in both directions. For every point in this grid, we compute
the combined signal and the background events 
to find the combination for which the required luminosity for a discovery ($\mc{L}_{\rm D}$) minimizes. We define $\mc{L}_{\rm D}$ as,
\be
\mc{L}_{\rm D} = {\rm Max}\lt(\mc{L}_5,\mc{L}_{10}\rt).\label{eq:ld}
\ee
Here, $\mc{L}_5$ is the luminosity required to attain a $5\sg$ statistical
significance for $\lt({\rm Sig.}/\sqrt{\rm Backg.}\rt)$ and $\mc{L}_{10}$ is the luminosity required to
observe 10 signal events. 
In Table~\ref{tab:cut_eff}, we display the `after-cut' cross sections of the combined signal and the dominant $Z+nj$ background (including the contributions from $ZV$) for the benchmark points of Table~\ref{tab:cut}. Though we show only the dominant background in the table, we include other sub-dominant contributions~\cite{Mandal:2012rx} (like inclusive top-pair etc.) while estimating $\mc L_{\rm D}$.

We show two $\mc{L}_{D}$ contours estimated for the 13 TeV LHC in Fig.~\ref{fig:LD}. To obtain this, we use constant $K_{\textrm{NLO}}=2$ for $M_{e_8}$ beyond 1.5 TeV like the recast analysis in section~\ref{sec:three}. With 300 fb$^{-1}$ of integrated luminosity, the mass reach goes from about 2.5 TeV to about 3.5 TeV as $\Lm$ decreases to about 3.5 TeV ($\Lm\approx M_{e_8}$) from very large values. Obviously, this increase in reach with decreasing $\Lm$ happens because of the single productions whose cross sections go like $1/\Lm^2$.

Before closing this section we make a note. Even though we have reinterpreted the CMS LQ data in 
terms of the $e_8$, it is also possible to separate them at the LHC. Let us suppose, a significant excess is found in the $eej$ data in future. In Fig.~\ref{fig:rapdist}, we show the $\et$
distribution of the second hardest electron (as an example), which  can be used to distinguish a spin-0 LQ from a spin-1/2 LG. Obviously, there are other possibilities as well. However, we do not pursue this issue further in this letter.

\section{Discussions and Conclusions}\label{sec:last}

The quark-lepton compositeness scenario is one of the well-known BSM scenarios which can accommodate LQs. In this letter, we have used the CMS first generation scalar LQ data in the $eejj$ and the $eej$ channels to probe this scenario. In these models, there exist other exotic composite particles that can also decay to lepton-jet final states. We have recast the CMS data in terms of such a particle,  the color octet partner of the SM electron. An $e_8$ decays to an electron and a gluon via a dimension five interaction, suppressed by the compositeness scale. This opens up the possibility of probing the compositeness scale with the $eejj$ and the $eej$ data.

In a recent  paper~\cite{Mandal:2015vfa}, we argued that at the LHC, a search for the pair production of a colored particle (generally, model independent) can get `contaminated' from the model dependent single productions and {\em vice versa}.
There, we used the examples of the CMS LQ searches to demonstrate how the pair and the single productions can be combined systematically in the signal simulations. As a result, even a search for the pair production can give information on the model parameters that control the single productions. In this letter too, we have adopted the same strategy, {\it i.e.}, we have recast both the $eejj$ and the $eej$ data with signals that are combinations of the pair and the single productions for different values of $\Lm$, the compositeness scale that controls the single productions. Hence, the analysis in this letter stands as yet another demonstration of our arguments in Ref.~\cite{Mandal:2015vfa}.

From the combined signal, we extract the exclusion limits in the $M_{e_8}$-$\Lm$ plane. The limits obtained by our analysis are not very precise as they are obtained by simple rescaling instead of a full statistical analysis. However, one can conclude that the $eejj$ data disfavor $e_8$'s with mass  below $\sim 1.5$ TeV for any value of $\Lm$.\footnote{If additional sources to the LG pair production (like the higher dimensional operators in footnote \ref{fn:footnote1} or the LO  electroweak gauge mediated pair production etc.) are considered, this limit would receive corrections and could acquire some $\Lm$-dependence even. However, it is normal to expect these corrections to be smaller than the QCD mediated LO pair production.} Beyond this mass range, the limit becomes a function of $\Lm$. 
As the mass increases, the single productions dominate the combined signals in both $eejj$ and $eej$ channels giving almost overlapping limits that can also be interpreted as the limits on $\Lm$. Data in both channels indicate that $\Lm\gtrsim 2$ TeV for 1.5 TeV $\lesssim M_{e_8}\lesssim$ 2 TeV. Beyond this mass range, where the exclusion limits enter in the region with $M_{e_8} > \Lm$, our effective theory approach becomes unreliable. We clearly mark this region in all the relevant plots. This is an inherent limitation present in any effective theory approach. It might also happen that, in nature, the  $e_8$ is actually heavier than the compositeness scale. In that case, all our limits/predictions would not be reliable except in the parameter region dominated by the (QCD mediated) pair production. For example, let us suppose that, in nature, $\Lm$ is actually smaller than 1.5 TeV, the mass range disfavored by the pair production data. In that case, we will still be able to say that the $e_8$ can not exist below 1.5 TeV but we would not be able to conclude anything definitively about $\Lm$ from our analysis. Notice that there are other higher dimensional operators (like $\mc O_{ggee}$ or $\mc O_{qqee}$ for contact interactions)  that, in principle, could also connect $\Lm$ with the $eejj/eej$ data irrespective of the values of $M_{e_8}$. However, two points go against them -- the first, the signal selection criteria are not designed to favor them, and the second, these operators are of dimensions higher than five (so unless $\Lm$ is very small, in which case the whole effective theory approach might break down, these terms are expected to be highly suppressed). Hence, despite the inherent limitation, our approach gives the best available limits on $\Lm$ and $M_{e_8}$ from the CMS 8TeV $eejj$ and $eej$ data within the domain of validity of the effective theory (compare the limit on $M_{e_8}$ with the limit quoted in the Particle Data Book~\cite{Agashe:2014kda}, $M_{e_8}>$ 86 GeV from old Tevatron data~\cite{Abe:1989es}).

Finally, we note that one can also analyze the second generation $\m\m jj/\m\m j$ data in terms of color octet muon. However, it will be a very similar exercise and we do not expect that it will provide very different limits on $\Lm$ than what we have obtained. In case of the LQ, production of the second generation is reduced compared to the first generation because of the relative suppression of the second generation quark PDFs. However, since the LG productions at the LHC are mainly gluon mediated, they remain roughly the same for any generation.

\begin{acknowledgments}
T.M. is supported by funding from the Carl Trygger Foundation under contract CTS-14:206 and the Swedish Research Council under contract 621-2011-5107.
\end{acknowledgments}


\begin{thebibliography}{100}



\bibitem{Pati:1974yy} 
  J.~C.~Pati and A.~Salam,
  \href{http://dx.doi.org/10.1103/PhysRevD.10.275}{Phys.\ Rev.\ D {\bf 10}, 275 (1974)}
  [\href{http://dx.doi.org/10.1103/PhysRevD.11.703.2}{Phys.\ Rev.\ D {\bf 11}, 703 (1975)}].
    
\bibitem{Terazawa:1976xx} 
  H.~Terazawa, K.~Akama and Y.~Chikashige,
  \href{http://dx.doi.org/10.1103/PhysRevD.15.480}{Phys.\ Rev.\ D {\bf 15}, 480 (1977)}.

\bibitem{Neeman:1979wp} 
  Y.~Ne'eman,
  \href{http://dx.doi.org/10.1016/0370-2693(79)90521-5}{Phys.\ Lett.\ B {\bf 81}, 190 (1979)}.
  
\bibitem{Harari:1979gi} 
  H.~Harari,
  \href{http://dx.doi.org/10.1016/0370-2693(79)90626-9}{Phys.\ Lett.\ B {\bf 86}, 83 (1979)}.
  
\bibitem{Shupe:1979fv} 
  M.~A.~Shupe,
  \href{http://dx.doi.org/10.1016/0370-2693(79)90627-0}{Phys.\ Lett.\ B {\bf 86}, 87 (1979)}.
  
\bibitem{Terazawa:1979pj} 
  H.~Terazawa,
  \href{http://dx.doi.org/10.1103/PhysRevD.22.184}{Phys.\ Rev.\ D {\bf 22}, 184 (1980)}.
    
\bibitem{Harari:1980ez} 
  H.~Harari and N.~Seiberg,
  \href{http://dx.doi.org/10.1016/0370-2693(81)90012-5}{Phys.\ Lett.\ B {\bf 98}, 269 (1981)}.
  
\bibitem{Fritzsch:1981zh} 
  H.~Fritzsch and G.~Mandelbaum,
  \href{http://dx.doi.org/10.1016/0370-2693(81)90626-2}{Phys.\ Lett.\ B {\bf 102}, 319 (1981)}.
  
\bibitem{Buchmuller:1986zs} 
  W.~Buchmuller, R.~Ruckl and D.~Wyler,
  \href{http://dx.doi.org/10.1016/0370-2693(87)90637-X}{Phys.\ Lett.\ B {\bf 191}, 442 (1987)}
  [Phys.\ Lett.\ B {\bf 448}, 320 (1999)].
    
\bibitem{Hewett:1997ce} 
  J.~L.~Hewett and T.~G.~Rizzo,
  \href{http://dx.doi.org/10.1103/PhysRevD.56.5709}{Phys.\ Rev.\ D {\bf 56}, 5709 (1997)}
  [hep-ph/9703337].
  
\bibitem{Kramer:1997hh} 
  M.~Kramer, T.~Plehn, M.~Spira and P.~M.~Zerwas,
  \href{http://dx.doi.org/10.1103/PhysRevLett.79.341}{Phys.\ Rev.\ Lett.\  {\bf 79}, 341 (1997)}
  [hep-ph/9704322].
  
\bibitem{Harari:1985cr} 
  H.~Harari,
  \href{http://dx.doi.org/10.1016/0370-2693(85)91518-7}{Phys.\ Lett.\ B {\bf 156}, 250 (1985)}.
  
\bibitem{Baur:1985ud} 
  U.~Baur and K.~H.~Streng,
  \href{http://dx.doi.org/10.1016/0370-2693(85)90946-3}{Phys.\ Lett.\ B {\bf 162}, 387 (1985)}.
  
\bibitem{Nir:1985ah} 
  Y.~Nir,
  \href{http://dx.doi.org/10.1016/0370-2693(85)90348-X}{Phys.\ Lett.\ B {\bf 164}, 395 (1985)}.
  
\bibitem{Rizzo:1985dn} 
  T.~G.~Rizzo,
  \href{http://dx.doi.org/10.1103/PhysRevD.33.1852}{Phys.\ Rev.\ D {\bf 33}, 1852 (1986)}.
  
\bibitem{Rizzo:1985ud} 
  T.~G.~Rizzo,
  \href{http://dx.doi.org/10.1103/PhysRevD.34.133}{Phys.\ Rev.\ D {\bf 34}, 133 (1986)}.
  
\bibitem{Streng:1986my} 
  K.~H.~Streng,
  \href{http://dx.doi.org/10.1007/BF01411142}{Z.\ Phys.\ C {\bf 33}, 247 (1986)}.
  
\bibitem{Khachatryan:2015vaa} 
  V.~Khachatryan {\it et al.} [CMS Collaboration],
  arXiv:1509.03744 [hep-ex].
 
\bibitem{Khachatryan:2015bsa} 
  V.~Khachatryan {\it et al.} [CMS Collaboration],
  \href{http://dx.doi.org/10.1007/JHEP07(2015)042}{JHEP {\bf 1507}, 042 (2015)}  
  [arXiv:1503.09049 [hep-ex]].
  
\bibitem{Aad:2015caa} 
  G.~Aad {\it et al.} [ATLAS Collaboration],
  \href{http://dx.doi.org/10.1140/epjc/s10052-015-3823-9}{Eur.\ Phys.\ J.\ C {\bf 76}, no. 1, 5 (2016)}
  [arXiv:1508.04735 [hep-ex]].
    
\bibitem{Khachatryan:2015qda} 
  V.~Khachatryan {\it et al.} [CMS Collaboration],
  arXiv:1509.03750 [hep-ex].

\bibitem{Celikel:1998dj} 
  A.~Celikel, M.~Kantar and S.~Sultansoy,
 \href{http://dx.doi.org/10.1016/S0370-2693(98)01299-4}{Phys.\ Lett.\ B {\bf 443}, 359 (1998)}.
  
\bibitem{Sahin:2010dd} 
  M.~Sahin, S.~Sultansoy and S.~Turkoz,
  \href{http://dx.doi.org/10.1016/j.physletb.2010.04.070}{Phys.\ Lett.\ B {\bf 689}, 172 (2010)}
  [arXiv:1001.4505 [hep-ph]].
  
\bibitem{Akay:2010sw} 
  A.~N.~Akay, H.~Karadeniz, M.~Sahin and S.~Sultansoy,
  \href{http://dx.doi.org/10.1209/0295-5075/95/31001}{Europhys.\ Lett.\  {\bf 95}, 31001 (2011)}
  [arXiv:1012.0189 [hep-ph]].
  
\bibitem{Jelinski:2015epa} 
  T.~Jeliński and D.~Zhuridov,
  \href{http://dx.doi.org/10.5506/APhysPolB.46.2185}{Acta Phys.\ Polon.\ B {\bf 46}, no. 11, 2185 (2015)}
  [arXiv:1510.04872 [hep-ph]].
  
\bibitem{Acar:2015wxp} 
  Y.~C.~Acar, U.~Kaya, B.~B.~Oner and S.~Sultansoy,
  arXiv:1511.05814 [hep-ph].

\bibitem{Chatrchyan:2012vza} 
  S.~Chatrchyan {\it et al.} [CMS Collaboration],
  \href{http://dx.doi.org/10.1103/PhysRevD.86.052013}{Phys.\ Rev.\ D {\bf 86}, 052013 (2012)}
  [arXiv:1207.5406 [hep-ex]].
    
\bibitem{Mandal:2012rx} 
  T.~Mandal and S.~Mitra,
  \href{http://dx.doi.org/10.1103/PhysRevD.87.095008}{Phys.\ Rev.\ D {\bf 87}, no. 9, 095008 (2013)}
  [arXiv:1211.6394 [hep-ph]].
    
\bibitem{Goncalves-Netto:2013nla} 
  D.~Goncalves-Netto, D.~Lopez-Val, K.~Mawatari, I.~Wigmore and T.~Plehn,
  \href{http://dx.doi.org/10.1103/PhysRevD.87.094023}{Phys.\ Rev.\ D {\bf 87}, 094023 (2013)}
  [arXiv:1303.0845 [hep-ph]].

\bibitem{Agashe:2014kda} 
  K.~A.~Olive {\it et al.} [Particle Data Group Collaboration],
  \href{http://dx.doi.org/10.1088/1674-1137/38/9/090001}{Chin.\ Phys.\ C {\bf 38}, 090001 (2014)}.

\bibitem{Mandal:2015vfa} 
  T.~Mandal, S.~Mitra and S.~Seth,
  \href{http://dx.doi.org/10.1007/JHEP07(2015)028}{JHEP {\bf 07}, 028 (2015)}
  [arXiv:1503.04689 [hep-ph]].

\bibitem{Alloul:2013bka} 
  A.~Alloul, N.~D.~Christensen, C.~Degrande, C.~Duhr and B.~Fuks,
  \href{http://dx.doi.org/10.1016/j.cpc.2014.04.012}{Comput.\ Phys.\ Commun.\  {\bf 185}, 2250 (2014)}
  [arXiv:1310.1921 [hep-ph]].

\bibitem{Pumplin:2002vw} 
  J.~Pumplin, D.~R.~Stump, J.~Huston, H.~L.~Lai, P.~M.~Nadolsky and W.~K.~Tung,
  \href{http://dx.doi.org/10.1088/1126-6708/2002/07/012}{JHEP {\bf 0207}, 012 (2002)}
  [hep-ph/0201195].
  
\bibitem{Alwall:2014hca} 
  J.~Alwall {\it et al.},
  \href{http://dx.doi.org/10.1007/JHEP07(2014)079}{JHEP {\bf 1407}, 079 (2014)}
  [arXiv:1405.0301 [hep-ph]].
 
\bibitem{Sjostrand:2006za} 
  T.~Sjostrand, S.~Mrenna and P.~Z.~Skands,
  \href{http://dx.doi.org/10.1088/1126-6708/2006/05/026}{JHEP {\bf 0605}, 026 (2006)}
  [hep-ph/0603175].
  
\bibitem{deFavereau:2013fsa} 
  J.~de Favereau {\it et al.} [DELPHES 3 Collaboration],
  \href{http://dx.doi.org/10.1007/JHEP02(2014)057}{JHEP {\bf 1402}, 057 (2014)}
  [arXiv:1307.6346 [hep-ex]].
  
\bibitem{Cacciari:2011ma} 
  M.~Cacciari, G.~P.~Salam and G.~Soyez,
  \href{http://dx.doi.org/10.1140/epjc/s10052-012-1896-2}{Eur.\ Phys.\ J.\ C {\bf 72}, 1896 (2012)}
  [arXiv:1111.6097 [hep-ph]].
  
\bibitem{Cacciari:2008gp} 
  M.~Cacciari, G.~P.~Salam and G.~Soyez,
  \href{http://dx.doi.org/10.1088/1126-6708/2008/04/063}{JHEP {\bf 0804}, 063 (2008)}
  [arXiv:0802.1189 [hep-ph]].
  
\bibitem{Alwall:2007fs} 
  J.~Alwall {\it et al.},
  \href{http://dx.doi.org/10.1140/epjc/s10052-007-0490-5}{Eur.\ Phys.\ J.\ C {\bf 53}, 473 (2008)}
  [arXiv:0706.2569 [hep-ph]].
  
\bibitem{Alwall:2008qv} 
  J.~Alwall, S.~de Visscher and F.~Maltoni,
  \href{http://dx.doi.org/10.1088/1126-6708/2009/02/017}{JHEP {\bf 0902}, 017 (2009)}
  [arXiv:0810.5350 [hep-ph]].
  
\bibitem{Abe:1989es} 
  F.~Abe {\it et al.}  [CDF Collaboration],
  \href{http://dx.doi.org/10.1103/PhysRevLett.63.1447}{Phys.\ Rev.\ Lett.\  {\bf 63}, 1447 (1989)}.
  
\end{thebibliography}
\end{document}